\documentstyle[12pt]{article}

\begin{document}

\title{U(1) Field Theory in the Limit of a Large Coupling Constant}

\author{Marco Frasca \\
Via E. Gattamelata,3 00176 Roma (Italia) \\
e-mail: marcofrasca@mclink.it
}

\date{17 June 1994}

\maketitle

\begin{abstract}
   We consider a massive fermion interacting with a U(1) gauge
   field in the limit of a large coupling constant. It is found that
   the current has a generalized London term that can originate
   massive excitations for two of the three components of the gauge field,
   which disappear for a free particle at rest.
   The origin of the superconductive term is due to a
   partial breaking of the gauge symmetry in the limit of a large coupling
   constant. Beside, the scalar potential generated by the particle,
   the only component of the gauge field that keeps gauge invariance,
   increases with the square of the distance.
   These results should give a path towards the derivation of quark
   confinement from QCD.
\end{abstract}

\section{Introduction}

The problem of extracting meaningful results from QCD, in the low-energy
limit or, better, when a large coupling is considered, is a fundamental
one
in particle physics. Small coupling perturbation theory gives results
in fine agreement with the high-energy behaviour of strong interactions
between quarks \cite{primo}. This is what one expects from an asymptotic free
theory that should describe that kind of physical interactions. In the
large coupling limit phenomenological theories exist without any
theoretical justification when QCD is taking in account. The foundations
of such theories are just plausibility arguments based on what one should
expect the full theory behaves in that limit \cite{secondo}. The other approach of
considering lattice QCD and solving the relative equations on a parallel
computer has, on the other side, to cope with a lattice length that in the
real world is zero and it is not so immediate to extrapolate the numerical
solutions to that limit \cite{terzo}. This scenario to treat QCD is founded on the
belief that only a small perturbation theory or, else, a perturbation
theory for a small development parameter, is possible in quantum
theory. Also asymptotic freedom is derived using results based on these
opinions and some authors cast doubts about it \cite{quarto}.

Actually, this is not all the thruth:
there exists a full theory for large perturbations in quantum
theory as I recently proved \cite{quinto}. That such a theory could be practically
applied for any time-dependent perturbation is showed in appendix to this
paper to make it self-consistent. The result is quite surprising for a
quantum
system with a large development parameter. Quite unexpected are also the
results we will show in this paper for a U(1) gauge theory giving a
possible
path to explain the quark confinement through QCD.

The guideline of this paper was put forward by T.D.Lee in Ref.\cite{sesto}:
``{\sl Quark confinement is a large-scale phenomenon.
Therefore, at least on the phenomenological level, it should be
understandable through a quasi-classical theory, much like the
London-Landau theory for superfluidity\ldots}''. The main points are
that the phenomenological level should be derived directly from the gauge
theory and that the equations for the current should contain
diamagnetic-like terms,
that is proportional to the vector potential. These we will prove for the
U(1)
gauge theory showing also that the scalar potential is a confining one,
that is proportional to a power of the distance in the limit of a large
coupling constant. The solution we will find
displays also a long-range excitation, this is
rensembling the Goldstone boson for the spontaneous breaking of symmetry.

The theory here considered is not fully quantized, that is we have no
second
quantization as this is not applicable in this case (we could not be able
to define a Berry's phase). A possible way out is the path integral
formulation
as given in \cite{settimo}. This question is not considered in the present paper as,
to obtain our results, the envisaged approach is enough and completes
the Lee's program, at least for the abelian theory.

Finally, our results display the high-energy behaviour of QED. The
meaning of them in such a case could be questioned as, at the
energy where our results could be worth, other interactions have to
be considered. Actually, the main aim of this paper is to show a path
toward a low-energy treatment of QCD.

The work is so structured. In section 2 we discuss the partial breaking
of symmetry in the strong coupling constant limit. In section 3 we derive
the U(1) current components in the same limit. In section 4 the case of
a free particle is discussed and the equations for the gauge field are
linearized showing massive excitation for two of the three components.
Finally, in section 5 we give the conclusions.

\section{The Breaking of the Gauge Symmetry in the Large Coupling
Limit}

A relativistic massive fermion is described by the Dirac equation that, by
requiring gauge invariance under U(1), can be cast in the form
\begin{equation}
    i\frac{\partial\psi}{\partial t} =          \label{eq:dirac}
    -i\mbox{\boldmath$\alpha$}\cdot\nabla\psi + \beta m\psi
    -q\mbox{\boldmath$\alpha$}\cdot{\bf A}\psi+qA_0\psi
\end{equation}
having considered the unit system with $\hbar=1$ and $c=1$. The coupling
constant $q$ is taken with the relative sign and the gauge field is
determined
through $\psi$. The case of interest is,
formally, the one with $|q|\rightarrow\infty$. From a physical point
of view, this represents the large coupling regime.
In order to cope with the sign dependence in the coupling constant, we
introduce the field $B_\mu$ defined through the equality $A_\mu = q
B_\mu$,
so that eq.(\ref{eq:dirac}) can be rewritten as
\begin{equation}
    i\frac{\partial\psi}{\partial t} =          \label{eq:dirac2}
    -i\mbox{\boldmath$\alpha$}\cdot\nabla\psi + \beta m\psi
    -q^2\mbox{\boldmath$\alpha$}\cdot{\bf B}\psi+q^2 B_0\psi
\end{equation}
that has the well-known form
\begin{equation}
    (H_0 + \lambda V)\psi = i\frac{\partial\psi}{\partial t}
\end{equation}
with $\lambda = q^2$,
$H_0 = -i\mbox{\boldmath$\alpha$}\cdot\nabla\psi + \beta m\psi$ and
$V = -\mbox{\boldmath$\alpha$}\cdot{\bf B} + B_0$.
In Refs.\cite{quinto}, I showed that a general perturbation series exists for this
equation, in the limit $\lambda\rightarrow\infty$, whose leading order has
the form
\begin{equation}
    V \psi = i\frac{\partial\psi}{\partial t}  \label{eq:order0}
\end{equation}
being the time variable rescaled as $t\rightarrow\lambda t$. Larger is the
coupling constant and more exact is eq.(\ref{eq:order0}). The problem
here, as pointed out in Refs.\cite{quinto}, is that we seem to have troubles when
$[V(t),V(t')]\neq 0$ being not able to solve that equation. Actually,
when the time scale is reset to its natural unit, the following
theorem holds (see the appendix for a proof),

\begin{em}
{\bf Theorem}:
   In the limit $\lambda\rightarrow\infty$, the solution of the equation
   \begin{equation}
       \lambda V(t) \psi = i\frac{\partial\psi}{\partial t}
\label{eq:schrod}
   \end{equation}
   can be cast in the form
   \begin{equation}
      |\psi> = \sum_n e^{-i\lambda\int_0^t dt' E_n(t')} a_n(t) |n;t>
   \end{equation}
   with
   \begin{equation}
      H(t) |n;t> = E_n(t) |n;t>,~~<m;t|n;t> = \delta_{mn}
   \end{equation}
   and
   \begin{equation}
      a_n(t) = e^{i\gamma_n(t)}a_n(0)
   \end{equation}
   with
   \begin{equation}
      \gamma_n(t) = \int_0^t dt' <n;t|i\frac{\partial}{\partial t}|n;t>.
   \end{equation}
\end{em}
It should be understood that
\begin{equation}
    a_n(0) =  <n;0|\psi(0)>.
\end{equation}
So, for large $\lambda$, we are not able to distinguish between a strongly
perturbed quantum system and an adiabatic one with the same initial
condition.

In our case we have
\begin{equation}
    q^2\left[-\mbox{\boldmath$\alpha$}\cdot{\bf B} + B_0\right]\psi =
    i\frac{\partial\psi}{\partial t}  \label{eq:scr}
\end{equation}
and we can conclude that, in the limit $|q|\rightarrow\infty$, the gauge
invariance is partially broken. In fact, the only transformations that
leave invariant the above equation are
\begin{equation}
    B_0\rightarrow B_0 - \stackrel{.}{\Lambda}({\bf x}, t)
\end{equation}
and
\begin{equation}
    \psi\rightarrow e^{iq^2\Lambda({\bf x}, t)}\psi.
\end{equation}
Then, gauge breaking terms are expected in the motion equations for ${\bf B}$,
that, in the London limit of a weak field, could give rise to massive
terms. For our definition of $B_\mu$, the relative motion equations are explicitly
independent on $q$, but could depend on it through $\psi$. We will see,
{\sl a posteriori}, that such a dependence is harmless, in the limit of a
strong coupling constant, being of oscillatory type.

\section{Derivation of the Current Components}

The aim of this section is to compute the eigenvalues and eigenvectors of
the equation
\begin{equation}
    \left[-q^2\mbox{\boldmath$\alpha$}\cdot{\bf B} + \label{eq:eigen}
           q^2 B_0 - E\right]u = 0
\end{equation}
so that, the solution of eq.(\ref{eq:scr}) can be written as
\begin{equation}
    \psi = N\sum_{i=1}^{4}a_i({\bf x})u_i({\bf x},t)  \label{eq:psi}
    \exp\left[-i\int_0^t dt' E_i({\bf x},t')+i\gamma({\bf x},t)\right]
\end{equation}
being $\gamma_i$ the geometrical phase of the i-fold state $u_i$
and the factor N is given by the condition
\begin{equation}
    \int_\Omega \psi^*\psi d^3x = 1.
\end{equation}
being $\Omega$ a normalization volume.
We have left the coefficient $a_i$ depend on ${\bf x}$ as this
is possible through the initial condition. We selected the following
eigenvectors for eq.(\ref{eq:eigen}):

For the eigenvalues $E_1=E_3 = q^2 (B_0 + |{\bf B}({\bf x},t)|)$
\begin{eqnarray}
    u_1 = \frac{1}{\sqrt{2}}\left(
          \begin{array}{c}
              -\frac{\mbox{\boldmath $\sigma$}\cdot {\bf B}}{|{\bf
B}|}\chi_1
              \\ \chi_1
          \end{array}\right)  &
    u_3 = \frac{1}{\sqrt{2}}\left(
          \begin{array}{c}
              -\frac{\mbox{\boldmath $\sigma$}\cdot {\bf B}}{|{\bf
B}|}\chi_2
              \\ \chi_2
          \end{array}\right);
\end{eqnarray}

For the eigenvalues $E_2=E_4 = q^2 (B_0 - |{\bf B}({\bf x},t)|)$
\begin{eqnarray}
    u_2 = \frac{1}{\sqrt{2}}\left(
          \begin{array}{c}
              \frac{\mbox{\boldmath $\sigma$}\cdot {\bf B}}{|{\bf
B}|}\chi_1
              \\ \chi_1
          \end{array}\right)  &
    u_4 = \frac{1}{\sqrt{2}}\left(
          \begin{array}{c}
              \frac{\mbox{\boldmath $\sigma$}\cdot {\bf B}}{|{\bf
B}|}\chi_2
              \\ \chi_2
          \end{array}\right)
\end{eqnarray}
having
\begin{equation}
    \chi_a^*\chi_b = \delta_{ab}.
\end{equation}
and
\begin{equation}
    \sigma_3\chi_a = (-1)^{a+1}\chi_a
\end{equation}
with $a,b=1,2$.
We see that there is not an explicit dependence on $|e|$ in the
eigenvectors,
as it has to be. The only dependence on the coupling constant in $\psi$
is through the exponentials. It is quite simple to verify that
\begin{equation}
    u_i^*u_j = \delta_{ij}
\end{equation}
and then, the normalization factor N is set by
\begin{equation}
   \frac{1}{N} = \sqrt{\int_\Omega \sum_{i=1}^{4}|a_i({\bf x})|^2 d^3x}.
\end{equation}
The geometrical phases are determined by the equation
\begin{equation}
    \gamma({\bf x}, t) = \frac{1}{2}\int_0^t dt'
    \frac{\stackrel{.}{B_2}({\bf x}, t')B_1({\bf x}, t')-
          B_2({\bf x}, t')\stackrel{.}{B_1}({\bf x}, t')}
         {|{\bf B}({\bf x}, t')|^2}
\end{equation}
that yields
\begin{eqnarray}
    \gamma_1 = i u_1^*\stackrel{.}u_1 =
    \gamma_2 = i u_2^*\stackrel{.}u_2 = \gamma \\ \nonumber
    \gamma_3 = i u_3^*\stackrel{.}u_3 =
    \gamma_4 = i u_4^*\stackrel{.}u_4 = -\gamma.
\end{eqnarray}

Now, we are in a position to compute the components of the current
that are given by
\begin{eqnarray}
    {\bf j} = \psi^* \mbox{\boldmath $\alpha$} \psi, & j_0 = \psi^*\psi
\end{eqnarray}
obtaining the following relations
\begin{equation}
    j_0 = N^2\sum_{i=1}^{4}|a_i({\bf x})|^2
\end{equation}
and
\begin{equation}
    j_k = N^2\sum_{i=1}^{4}(-1)^i|a_i({\bf x})|^2\frac{B_k}{|{\bf B}|} +
          j^{(ND)}_k
\end{equation}
where ND stays for Not Diagonal having
\begin{eqnarray}
    j^{(ND)}_k = N^2 i\epsilon_{ijk}\frac{B_j}{|{\bf B}|}\left[
    a_1^*a_2 \chi_1^*\sigma_k\chi_1 e^{2i\beta({\bf x},t)} + \right. \\
    a_1^*a_4 \chi_1^*\sigma_k\chi_2
           e^{2i\beta({\bf x},t) - 2i\gamma({\bf x},t)} +      \\
\nonumber
    a_3^*a_2 \chi_2^*\sigma_k\chi_1
           e^{2i\beta({\bf x},t) - 2i\gamma({\bf x},t)} +      \\
\nonumber
    a_3^*a_4 \chi_2^*\sigma_k\chi_2\left.
           e^{2i\beta({\bf x},t)} - c.c.\right].               \\
\nonumber
\end{eqnarray}
being $\beta({\bf x},t) = \int_0^t dt' |{\bf B}({\bf x},t')|$.
So, we have a general kind of diamagnetic relation between the field and
the current that can be cast in the form (repeated indeces mean summation)
\begin{equation}
    j_k = \rho_{km}({\bf x}, t)\frac{B_m}{|{\bf B}|}
\end{equation}
and the tensor $\rho_{km}$ is determined just by the initial state of the
particle and that of the gauge field.
In the next section we will show, in a particular
case, as, due the above relation between current and field,
massive excitations of the field can arise.

\section{The Case of a Negative Charged Free Particle}

For a negative charged free particle we take
\begin{equation}
    \psi({\bf x},0) =
    \left(\frac{\epsilon_p + m}{2\epsilon_p}\right)^\frac{1}{2}
    \left(
    \begin{array}{c}
        {\displaystyle -\frac{p}{\epsilon_p + m}}\chi_1 \\
              \chi_1
    \end{array}
    \right)
    e^{ipz}
\end{equation}
representing a particle moving along direction $z$ with momentum $p$ and
energy $\epsilon_p = \sqrt{p^2+m^2}$. Taking
\begin{eqnarray}
    B_1({\bf x},0) = B_2({\bf x},0) = 0, &
    B_3({\bf x},0) = \Delta = const > 0
\end{eqnarray}
we get
\begin{eqnarray}
    a_1 = \frac{N}{2}
          \left(\frac{\epsilon_p + m}{\epsilon_p}\right)^\frac{1}{2}
          \left(1 + \frac{p}{\epsilon_p + m}\right)e^{ipz} \\ \nonumber
    a_2 = \frac{N}{2}
          \left(\frac{\epsilon_p + m}{\epsilon_p}\right)^\frac{1}{2}
          \left(1 - \frac{p}{\epsilon_p + m}\right)e^{ipz} \\
   a_3 = a_4 = 0. \nonumber
\end{eqnarray}
It is a simple matter to find $N = \frac{1}{\sqrt{\Omega}}$. So, at last,
\begin{equation}
    \rho_{ik} = -\frac{p}{\epsilon_p}\frac{1}{\Omega}\delta_{ik} -
                     \frac{m}{\epsilon_p}\frac{1}{\Omega}
                     \epsilon_{ijk}\chi_1^*\sigma_k\chi_1\sin(2\beta).
\end{equation}
The first term on the rhs disappears for a particle at rest. But we
are interested to the high-energy case where
the last term can be neglected and then we are left
with the following equations for the components of the gauge field
\begin{eqnarray}
    \Box {\bf A} - \nabla(\nabla\cdot{\bf A}) = \eta^2
    \frac{\bf A}{|{\bf A}|}\label{eq:field} \\
    \triangle_2 A_0 = \eta^2 \nonumber
\end{eqnarray}
with ${\displaystyle \eta^2 = \frac{|q|}{\Omega}}$
(being ${\displaystyle \frac{p}{\epsilon_p}\approx 1}$),
and use has been made of the residual gauge freedom
to make $A_0$ time-independent.

In order to see if eqs.(\ref{eq:field}) admit ``massive'' solutions,
we try to linearize around the value $\Delta$,
we put
\begin{eqnarray}
    A_1 = \delta A_1          \\  \nonumber
    A_2 = \delta A_2          \\
    A_3 = \Delta + \delta A_3     \nonumber
\end{eqnarray}
so that
\begin{equation}
    \frac{1}{|{\bf A}|} \approx \frac{1}{|\Delta|}\left(1 -
    \frac{\delta A_3}{\Delta}\right)
\end{equation}
yielding the linearized equations
\begin{eqnarray}
    \Box \delta A_1 - \partial_1(\nabla\cdot\delta{\bf A}) =
    \mu^2 \delta A_1 \\
    \Box \delta A_2 - \partial_2(\nabla\cdot\delta{\bf A}) =
    \mu^2 \delta A_2 \\
    \Box \delta A_3 - \partial_3(\nabla\cdot\delta{\bf A}) =
    \eta^2.                                            \label{eq:dA3}
\end{eqnarray}
where set ${\displaystyle \mu^2 = \frac{\eta^2}{\Delta}}$.
Then, in the limit of weak field, the components $A_1$ and $A_2$ acquires a
mass term in their motion equations. However, this approximation, at least
for the component $\delta A_3$, ceases to be valid for times of the order of
$\mu^{-1}$. This is due to the existence of
a part of the solution increasing with the square of time originating from
the constant on the rhs of eq.(\ref{eq:dA3}). Anyway, we should keep in mind
that the correct equations are (\ref{eq:field}). The $A_3$ component of the
linearized gauge-field remembers very nearly the Goldstone boson of the
spontaneous breaking of symmetry.

The meaning of the constant $\Delta$ is quite clear and assumes the same
role of the vacuum value of a Goldstone field. Simply, we have that
initially there is no field and then, we can assume it to be
$\nabla\chi({\bf x})$, with $\chi({\bf x})$
something like $constant \cdot z$. We have taken this constant to be
positive.
Actually, the sign plays no role, we have chosen that one for simplify
formulas.
More complex gauges could be selected at a cost to loose contact with the
London limit in the weak field approximation.

As a last consideration, we observe that the component $A_0$ goes like
$r^2$, that is we have a confining potential. This result appears very
encouraging in view of the application of the present theory to QCD.

\section{Conclusions}

We showed that, in the large coupling limit of the U(1) gauge theory,
the gauge symmetry is partially broken giving rise to possible mass
terms in the London limit of weak field, and to a confining potential
for the fourth component $A_0$. The case we considered is that of
a free negative charged particle. In the same approximation, we
showed that a field component has long-range effect for its time of
validity,
very similarly to the Goldstone boson in the spontaneous breaking of a
symmetry. The diamagnetic-like behaviour of the abelian gauge field,
in the limit of a strong coupling constant,
could give a path toward the explanation of the quark confinement
through QCD, where the considered limit is a more natural one.

\section*{Appendix}

In order to proof our theorem we seek the solution of eq.(\ref{eq:schrod})
in the form
\begin{equation}
   |\psi> = \sum_n a_n(t) e^{-i\lambda\int_0^t dt' E_n(t')} |n;t>.
\end{equation}
A direct substitution gives
\begin{equation}
   i\frac{da_m}{dt} = -\sum_n e^{-i\lambda\int_0^t dt' [E_n(t') -
E_m(t')]}
                      <m;t|i\frac{\partial}{\partial t}|n;t> a_n(t).
\end{equation}
If we set
\begin{equation}
   a_m(t) = b_m(t) e^{i\gamma_m(t)},
\end{equation}
being
\begin{equation}
   \gamma_m(t) = \int_0^t dt' <m;t'|i\frac{\partial}{\partial t'}|m;t'>
\end{equation}
the geometrical phase, it easily obtained, for the generic $b_m$, the
equation
\begin{equation}
   i\frac{db_m}{dt} = -\sum_{n\neq m}
                      e^{-i\lambda\int_0^t dt' [E_n(t') - E_m(t')]}
                      e^{i[\gamma_n(t) - \gamma_m(t)]}
                      <m;t|i\frac{\partial}{\partial t}|n;t> b_n(t).
\end{equation}
Now we observe that the term
\begin{equation}
   g_{mn}(t) = e^{i[\gamma_n(t) - \gamma_m(t)]}
               <m;t|i\frac{\partial}{\partial t}|n;t>
\end{equation}
does not depend on $\lambda$, while $\lim_{\lambda\rightarrow\infty}
b_m(t)$
must be finite in order to be $\sum_n |b_n(t)|^2 = 1$. So, writing
\begin{equation}
   b_m(t) = b_m(0) + i\sum_{n\neq m}\int_0^t dt'
            e^{-i\lambda\int_0^{t'} dt'' [E_n(t'') - E_m(t'')]}
            g_{mn}(t') b_n(t'),        \label{eq:stat}
\end{equation}
by invoking the stationary phase method \cite{ottavo}, it is immediate to realize
that
\begin{equation}
   \lim_{\lambda\rightarrow\infty} b_m(t) = b_m(0) = a_m(0)
\end{equation}
that proves the theorem. It easy to see that no adiabatic hypothesis is
entered in our proof. It should be noted that we used the stationary phase
method that implies that our method can only be applied to time-varying
problems. In fact, for a time-independent case, we would obtain that
the convergence of the integral in eq.(\ref{eq:stat}) cannot be assured
for the point at infinity.

It should be kept in mind that this theorem gives also an asymptotic
approximation to the solution of eq.(\ref{eq:schrod}) for $\lambda$
large but finite. In fact, the results of Refs.\cite{quinto} can be easily
recovered with this approach. Then, we conclude that, in quantum
mechanics,
strongly perturbed quantum system are not distinguishable from adiabatic
ones with the same initial conditions, as greater is the perturbation.
Else,
we can say that the adiabatic approximation is a good asymptotic
approximation
for quantum system with a large coupling constant $\lambda$.

\end{document}